\pdfoutput=1

\documentclass[11pt]{article}

\usepackage[]{ACL2023}

\usepackage{times}
\usepackage{latexsym}

\usepackage[T1]{fontenc}

\usepackage[utf8]{inputenc}

\usepackage{microtype}

\usepackage{amsmath,amssymb,amsfonts}
\usepackage{algorithmic}
\usepackage{graphicx}

\usepackage{caption}
\usepackage{subcaption}
\usepackage{tabularx}
\usepackage{xspace}
\usepackage{paralist}

\usepackage{color-edits}
\addauthor{gb}{red}
\addauthor{adam}{green}

\newcommand{\gagan}[1]{{\textcolor{purple}{[[GB: #1]]}}}

\newcommand{\ignore}[1]{}

\usepackage{array}
\newcolumntype{l}{>{\centering\arraybackslash}m{3cm}}

\usepackage{fontawesome}

\newcommand{\passOne}{\textsc{Pass}\xspace}
\newcommand{\editSim}{\textsc{Edit-Sim}\xspace}
\newcommand{\combined}{\textbf{\textsc{Combined}}\xspace}
\newcommand{\bleu}{\textsc{Bleu}\xspace}
\newcommand{\perValue}{\textsc{Value}\xspace}
\newcommand{\perAccuracy}{\textbf{\textsc{Accuracy}}\xspace}
\newcommand{\perEffort}{\textbf{\textsc{Effort}}\xspace}

\title{Aligning Offline Metrics and Human Judgments of\\ Value for Code Generation Models}

\author{Victor Dibia\textsuperscript{1},
  Adam Fourney\textsuperscript{1},
  Gagan Bansal\textsuperscript{1}, \\
  \textbf{Forough Poursabzi-Sangdeh\textsuperscript{1},
  Han Liu\textsuperscript{2},
  Saleema Amershi\textsuperscript{1} }\\
  \textsuperscript{1}Microsoft Research, Redmond, United States \\
  \texttt{\{victordibia, adam.fourney, gaganbansal, } \\
   \texttt{\ fpoursabzi, samershi\}@microsoft.com, \texttt{hanliu@uchicago.edu}} \\
  \textsuperscript{2}University of Chicago, Chicago, United States \\
}

\begin{document}
\maketitle
\begin{abstract}

Large language models have demonstrated great potential to assist programmers in generating code. For 
such human-AI pair programming scenarios,
we empirically demonstrate that while generated code are most often evaluated in terms of their {\em functional correctness} (i.e., whether generations pass available unit tests), correctness does not fully capture (e.g., may underestimate) the {\em productivity} gains these models may provide. Through a user study with $N=49$ experienced programmers, we show that while correctness captures high-value generations, programmers still rate code that fails unit tests as valuable if it reduces the overall effort needed to complete a coding task. Finally, we propose a hybrid metric that combines functional correctness and syntactic similarity and show that it achieves a 14\% stronger correlation with value and can therefore better represent real-world gains when evaluating and comparing models. 

\end{abstract}

\section{Introduction}

Large language models trained on code (e.g., Codex \cite{chen2021evaluatingcodex}, AlphaCode \cite{li2022competitionalphacode}, CodeGen \cite{nijkamp2022conversationalcodegen}, InCoder \cite{fried2022incoder}) have shown impressive capabilities on code generation tasks. One important application for such models is {\em Human-AI pair programming}, where a model suggests in-line code completions (e.g., within an IDE) that programmers can choose to ignore, accept, or edit as needed. Early studies suggest that this paradigm may dramatically boost productivity and transform the practice of software development \cite{ziegler2022productivity,github2022blog}.

\begin{figure}
  \centering
  \includegraphics[width=\columnwidth]{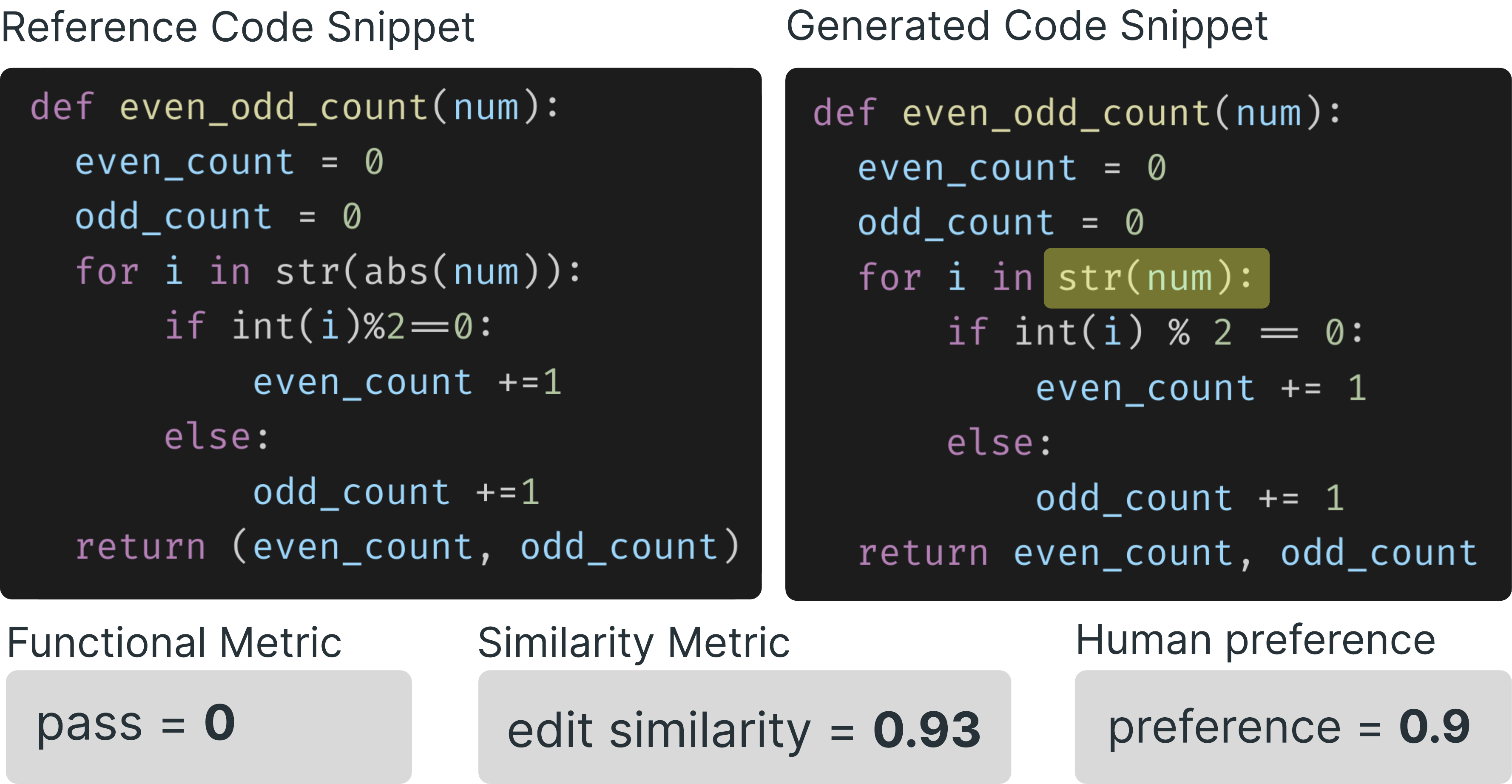}
  \caption{In the example above (counting even and odd numbers), code suggested by a  model fails unit tests but is deemed useful by programmers because adding a short check ($abs$ value) fixes the generation.
  \label{fig:studybanner}} 
\end{figure}
 
As is common with model development more generally, code-generation advances are largely driven by comparing model performance on {\em offline metrics} (i.e., metrics computed automatically over held out evaluation data) that can be easily tracked on leaderboards. 
{\em Functional correctness} metrics such as \textit{pass@k} \cite{chen2021evaluatingcodex} currently represent the state-of-best-practice~\cite{chen2021evaluatingcodex,fried2022incoder,austin2021programgoogle,chowdhery2022palm,nijkamp2022conversationalcodegen,hendrycks2021measuring, kulal2019spoc}. These metrics evaluate generations by executing a set of unit tests and assessing whether the generations pass or fail. 
While functional correctness is clearly important, it does not fully capture the productivity gains programmers may value about code generation assistance. For example, a generation that fails unit tests might yet provide critical hints to solve a task (see example in Fig \ref{fig:studybanner}), or serves as boilerplate that can be adapted with minimal effort. Likewise, functionally correct code might be difficult to read or maintain, or may contain other vulnerabilities.

With developer productivity in mind ~\cite{forsgren2021the}, we investigate syntactic similarity-based offline performance metrics (e.g., \cite{svyatkovskiy2020intellicode, chowdhery2022palm, papineni2002bleu}) as proxies of programmer effort needed to modify or correct automatic code generations. Similarity-based metrics compute how similar a generation is to reference or ground truth code, typically available in the offline setting. We then conducted a user study with N=49 experienced programmers to assess how well self-reported utility ~\cite{forsgren2021the} correlates with similarity-based and functional correctness metrics. Our work answers the following key research questions:
\begin{compactenum}
\item Do programmers still value code generations that may be incorrect (fail unit tests)?
\item How well do existing offline performance metrics align with programmer-rated value, accuracy and effort?
\item Does a metric that captures both functional correctness and effort saved better align with programmers' perceived value?
\end{compactenum}


 In our studies, we showed participants code generated by AI models and asked them to provide ratings in terms of the accuracy of the code, overall value of the code and effort associated with fixing the code (if any). We find that while ratings on effort and accuracy both correlate with value, effort is significantly more correlated. In other words, code that is perceived as easy-to-fix is judged to be more valuable. Conversely, when considering offline metrics, we find that while functional correctness metrics are more correlated to value compared to similarity based metrics, similarity based metrics offer complementary information. Specifically, we find 42\% of generations that failed unit tests were still rated as valuable - and similarity based metrics provide a better signal as to value in this regime.  We therefore propose a metric that combines functional correctness and similarity and show that it increases correlation with perceived value by 14\%.

\section{Related Work}

Offline performance evaluation of AI models typically consists of running models as isolated components over benchmark datasets and then computing aggregate \textit{metrics} (e.g., accuracy, AUC, and precision/recall) that can be easily compared and tracked on leaderboards. While these evaluation practices have led to rapid advancements in AI by enabling efficient apples-to-apples model comparison, a growing body of work has raised concerns about the mismatch between popular metrics and what people need and value in the real world \cite{THOMAS2022100476, Raji_2022, hellendoorn2019,Hand_2006, Jacobs_2021, neurips-2020-tutorial:beyond-accuracy, zhou2022deconstructing}. 
Using metrics that fail to appropriately capture what people value can result in deploying models that are at best less effective than they could be, and at worst harmful to people and society \cite{THOMAS2022100476, Raji_2022, Hand_2006}. In this work, we investigate the extent to which common offline code generation metrics capture what professional programmers value about code generation models. In particular, we examine how well existing code generation metrics capture notions of developer effort and productivity \citet{forsgren2021the}.



The current most popular family of code generation metrics is based on measuring \textit{functional correctness}. 
Functional correctness metrics seek to evaluate generated code against known objective properties such as passing unit tests \cite{chen2021evaluatingcodex, austin2021programgoogle, li2022competitionalphacode, roziere2020unsupervisedtranlate}. 
Following the release of Codex and the HumanEval dataset \cite{chen2021evaluatingcodex}---which is a dataset of 164 hand-written problems in python with associated unit tests---the functional correctness metric of $pass@k$ (where $k$ code samples are generated per problem and a problem is considered solved if any of the $k$ generations passes the corresponding unit tests) has emerged as the dominant method for evaluating code generation models (e.g., \cite{fried2022incoder, xu2022systematicpolycoder, li2022competitionalphacode,austin2021programgoogle}). Advocates of functional correctness metrics argue for their resemblance to programming best practices (e.g, test-driven development) and fidelity to capturing functional behaviour \cite{chen2021evaluatingcodex}. However, in this work we demonstrate that functional correctness does not fully capture what people value about code generation models.


Similarity-based metrics compare tokens from generated code to tokens of known solutions, with code that is more similar to given solution(s) being considered better. Multiple similarity-based metrics have been proposed for evaluating code generation models including 
exact match \cite{lu2021codexglue}, edit distance \cite{svyatkovskiy2020intellicode, chowdhery2022palm}, BLEU \cite{papineni2002bleu}, CodeBLEU \cite{ren2020codebleu}, and ROGUE \cite{lin2004rouge}.
Analyses of similarity-based metrics and other measures of code quality have been mixed (e.g., \cite{ren2020codebleu} vs \citet{austin2021programgoogle}). However, in most of these cases, similarity was considered a proxy for functional correctness. In this work, we revisit similarity-based metrics as proxies for effort saved in coding tasks \citet{svyatkovskiy2020intellicode} and demonstrate how they can be used to better capture value.



In this work we focus on $pass@k$ as a proxy for functional correctness and we experiment with two similarity-based metrics, namely, normalized edit similarity \cite{lu2021codexglue, svyatkovskiy2020intellicode, chowdhery2022palm} (which measures how many single-character edits--- including insertion, substitution, or deletion---are required to convert generated code to some reference code) and BLEU (which measures the token overlap between the generated and reference text) to investigate how these metrics approximate different facets of what programmers value in practice.
  
\section{User Study}

We designed a user study to evaluate how well functional correctness- and similarity-based offline metrics approximate value of code generations for programmers. The study showed experienced programmers various programming tasks, together with code generations and reference solutions. Programmers then rated the generations on perceived accuracy, effort, and value.

\subsection{Dataset for Programming Tasks}
We selected programming tasks from the HumanEval dataset \cite{chen2021evaluatingcodex}, which consists of 164 hand-crafted programming tasks and solutions written in Python. Each task includes a task description (i.e., a function header followed by a comment describing the task with some sample test cases (115 – 1360 characters)), a canonical hand-written solution, and a set of associated unit tests. HumanEval has been extensively used to evaluate code generation systems (e.g., \cite{chen2021evaluatingcodex, fried2022incoder, xu2022systematicpolycoder, chowdhery2022palm, nijkamp2022conversationalcodegen}). 
To the best of our knowledge, HumanEval is not part of any model's training data, and its simple standalone tasks makes it an ideal choice for user studies.

\subsection{Offline Metrics for Code Generation}
We experimented with three offline metrics, one of which served as a proxy for functional correctness and the other two served as a proxy for a programmer's effort.

\paragraph{\passOne:} As a proxy for functional correctness, we computed the $pass@k$ metric \cite{chen2021evaluatingcodex}. $pass@k$ takes $k$ generations for a problem and considers the problem solved if any generation passes the accompanying unit tests (in our case the unit tests provided in the HumanEval dataset). While related work has presented $pass@k$ results for values of $k$ including 1, 10, and even up to 1M \cite{chen2021evaluatingcodex,li2022competitionalphacode}, we focus our analysis on $k=1$ which most closely resembles the real-world scenario where a programmer sees a single generation inline within a coding editor. 

\paragraph{\editSim:} As one proxy for effort, we computed normalized edit similarity \cite{svyatkovskiy2020intellicode} as follows:

\[ \editSim = 1 - \dfrac{lev(gen, ref)}{max(len(gen), len(ref) )}
\]

where ${gen}$ is code generated by a model for a problem in the HumanEval dataset,  ${ref}$ is the hand-written reference solution to the problem and ${lev}$ is the character Levenshtein edit distance.
    
\paragraph{BLEU:} As another proxy for effort, we computed BLEU using the formulation introduced by \citet{papineni2002bleu} (generated code compared with a single reference), and based on the implementation in the Tensorflow library \cite{tensorflow2015-whitepaper}.

We focused on syntactic similarity-based metrics like \editSim \citet{lu2021codexglue,svyatkovskiy2020intellicode, chowdhery2022palm} and BLEU \citet{barone2017parallel, karaivanov2014phrase, nguyen2013lexical,ahmad2021unified,wang2021codet5} because they have been commonly-used metrics for  evaluating text-based generative models, especially, for code generation scenarios.

\subsection{Code Generation Models} \label{sec:studymodels}
We selected 5 publicly available autoregressive large language models trained on code, varied mostly by the parameter size of each model. The first two models are  variants of the CodeGen model introduced by \citet{nijkamp2022conversationalcodegen} (\href{https://huggingface.co/Salesforce/codegen-350M-multi}{CodeGen350 Multi}, \href{https://huggingface.co/Salesforce/codegen-2B-multi}{CodeGen2B Multi})  - autoregressive transformers with the regular next-token prediction language modeling as the learning objective trained on a natural language corpus and programming language (C, C++, Go, Java, JavaScript, and Python) data curated from GitHub. Next, we use three publicly available variants of the \href{https://beta.openai.com/docs/models/codex}{Codex model}   \cite{chen2021evaluatingcodex}, a GPT language model fine-tuned on publicly available code from GitHub (Cushman, Davinci1, Davinci2).  Note that the goal of this work is to compare code-generation {\em metrics} and not to assess the performance of models. We used models of different sizes to help ensure our findings on how metrics behave translate across a range of model qualities. Following guidance from \citet{chen2021evaluatingcodex} who demonstrate the importance of optimizing sampling temperature for particular values of $k$, we used a low temperature value of $t = 0.2$ for $k=1$ so that each model generates the most likely code tokens.

\subsection{Tasks} 


We used programming tasks from the HumanEval dataset, where for each task, participants were shown the task description (function header and docstring describing the task along with sample test cases), the corresponding unit tests, and two code snippets  --- the reference solution from the HumanEval dataset and a generation for that task from one of the models --- shown in a random order. Each snippet was randomly assigned a name - \textit{Code Snippet A} or \textit{Code Snippet B} for easy reference in the subsequent questions. All parts of the interface showing code were syntax highlighted to improve readability.

For each task, participants answered questions designed to collect their judgements along three dimensions of interest: overall value, accuracy, and effort which we hypothesized would impact value. Each question used 5-point Likert scales and were shown sequentially only after the previous question had been answered. The questions were as follows:

\paragraph{\perAccuracy:} The first question asked participants to judge whether both snippets were functionally equivalent. Since the reference solution is correct, functional equivalence can be used to infer perceived accuracy of a generation (complete equivalence indicates the participant believes the generation would produce the same outputs for all the same inputs as the reference solution which passes the provided unit tests). 
 We used this equivalence question-based approach to assess perceived accuracy because our pilots suggested that judging equivalence is easier than solving the coding task from scratch, and also because it enabled us to design a simpler, consistent survey -- the other two survey questions (as described next) also compared the generation to the reference.
     
At this point in the task, participants were not told which snippet corresponded to the generation and which was written by a human programmer to minimize the impact of any existing biases about the capabilities of AI models.  

\paragraph{\perValue:} Once participants advanced to the second question, the interface disclosed which snippet (A or B) was AI generated and which was a reference solution. They were then asked how useful the generated snippet would be assuming they were a programmer attempting to solve the task themselves. We described usefulness in terms of whether participants believed the generation provided a useful starting point, ranging from Extremely useful (they "would definitely accept it and use it as a starting point") to Not at all useful (they "would not even want to see the generation" let alone accept and use it as a starting point). 

\paragraph{\perEffort:} The final question asked participants how much effort they believed it would require them to modify the AI generated solution into a correct solution similar to the snippet written by a human programmer, if any.


\subsection{Study Protocol and Participants}  
The study consisted of four sections: consent form, instructions, main study, and a brief post-study feedback section. The instructions section was a sample task designed to familiarize participants with the mechanics of the study interface (e.g., they will be shown problems and asked to provide ratings, they will not be allowed to go back and revise previous responses) and to anchor them to pair programming scenario. 

The main study was made up of 12 tasks.
We chose 12 because our pilot studies showed that participants could complete 12 tasks within an hour. For each task, participants were shown a generation from one randomly chosen model from our set of 5 models.

A key goal of our study was to assess how well our offline metrics of interest align with what programmers value. We were particularly interested in understanding the tradeoffs between functional correctness and similarity as they relate to value and so we wanted to probe cases where these metrics disagreed. Therefore, to select study tasks, we first considered taking a random sample from HumanEval. However, the number of generations falling into regions where these metrics agreed on the largest model (Davinci2) was over-represented compared to the disagreement region (70\% agreement vs 30\% disagreement). Therefore, we chose a stratified sampling method where we first assigned each HumanEval problem into one of three buckets: \passOne = 1 and \editSim is low,  \passOne = 0 and \editSim is high, \passOne and \editSim agree
 .\footnote{According to Davinci2 and where similarity was thresholded along the median similarity value for that model.} Then, we sampled equally across each bucket aiming to annotate 20 problems per bucket for this study.

Because we intended to recruit professional programmers, we aimed to obtain up to 2 annotations per problem-model pair. With 60 problems (20 per bucket), 5 models, and 2 annotations per task and a budget of 12 problems per participant, this required us to recruit 50 participants for this study. We assigned annotation tasks to participants by randomly sampling a problem from our sample of 60 and then randomly sampling a generation for that problem, without repeating a problem for any participant, until each problem-model pair was assigned 2 annotations. 

We recruited professional programmers from a large technology company for this study and recruitment emails were sent out to a randomly sampled subset of software engineers. Participants were required to have at least 1-2 years of programming experience with Python and to program in Python at least a few times a year. 61\% of respondents indicated they had worked on a python project in the last month and 59\% had never used a pair programming AI assistant like GitHub Copilot.

The study was deployed as a web application. Participants were given five days to complete the study, and could pause and resume using their personalized study link. At the end of the study, participants were given a \$50 online gift card. As an additional incentive, we awarded the top 5 performers an additional \$50 gift card. We determined top performers based on the rate at which participants correctly indicated a generation was equivalent to the reference code when it passed vs when it failed the given unit tests. This experiment was approved by our organization's internal IRB process.

\section{Study Results}




\begin{figure}
  \centering
  \includegraphics[width=\columnwidth]{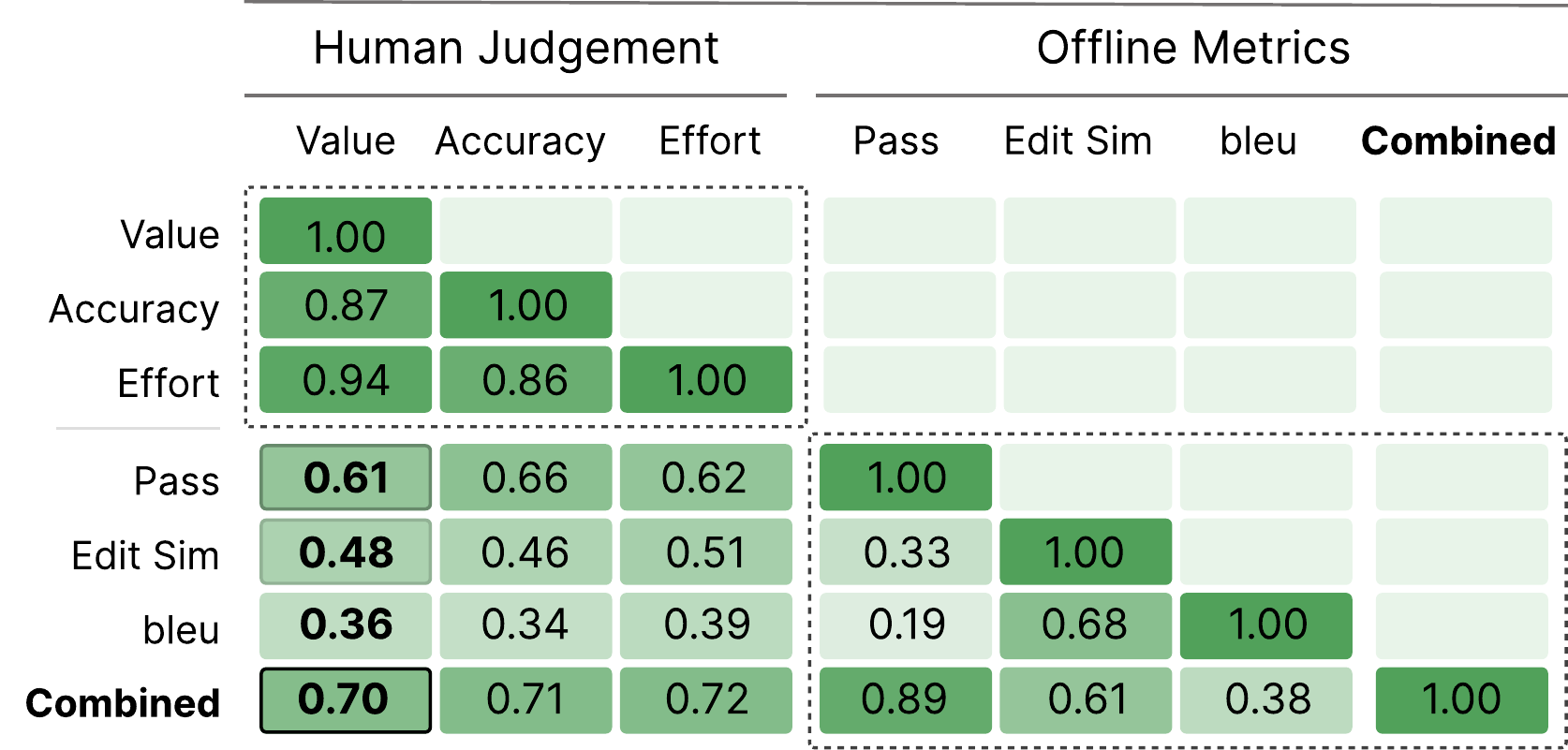}
  \caption{Correlation (Pearson) between human judgements (perceived value, accuracy and effort) and offline metrics (functional correctness, edit similarity and a combined metric (see section \ref{sec:combinedmetric})). All correlations are significant with $p < 0.001$. 
  \label{fig:studyheader}} 
\end{figure}
At the end of the study period, we obtained responses from 49 participants. We then applied the following criteria to evaluate the quality of responses: First, we computed the median response time per task for all participants and also computed a performance rating on the code equivalence task in the same way we determined top performers in our study. Data from three participants who fell within the bottom 10th percentile of the median task completion times and their performance was worse than the probability of random chance (given the questions they responded to) was excluded from the data analysis. The final dataset includes data from 46 participants with 552 annotations across 290 unique tasks and 1.96 annotation per task. Finally, across tasks where we obtained multiple annotations, we verified that there was agreement between annotators\footnote{In 50\% of cases, annotators are in perfect agreement; 75\% differ by at most one point in valence (on a rating scale of 1-5) and the mean difference is 0.89.} and then computed the mean annotation per task for use in our analysis. In this section, we present the main findings based on this data.

\subsection{Accuracy is Valuable, but Effort Matters}
Our first finding is that the \perValue of a generation is nearly perfectly correlated with the perceived \perEffort needed to correct a generation (Pearson $r = 0.94$; 95\%-confidence interval [$0.92-0.95$]). Recall that \perEffort is reverse-coded such that a score of 5 indicates ``no effort'' is needed. \perAccuracy is also highly correlated (Pearson $r = 0.87$; 95\%-confidence interval [$0.84-0.90$]), but significantly less so -- we note that their confidence intervals do not overlap. \textbf{From this we conclude that \perAccuracy isn't everything, and \perEffort is at least as important a signal for capturing \perValue.} We will return to this theme throughout the paper. Correlations between these dimensions are presented in the top-left quadrant of Figure \ref{fig:studyheader}.

\subsection{Offline Metrics Highly Correlate with Programmers' Judgements, But There is Room for Improvement}

\begin{figure}[t]
  \centering
  \includegraphics[width=.85\columnwidth]{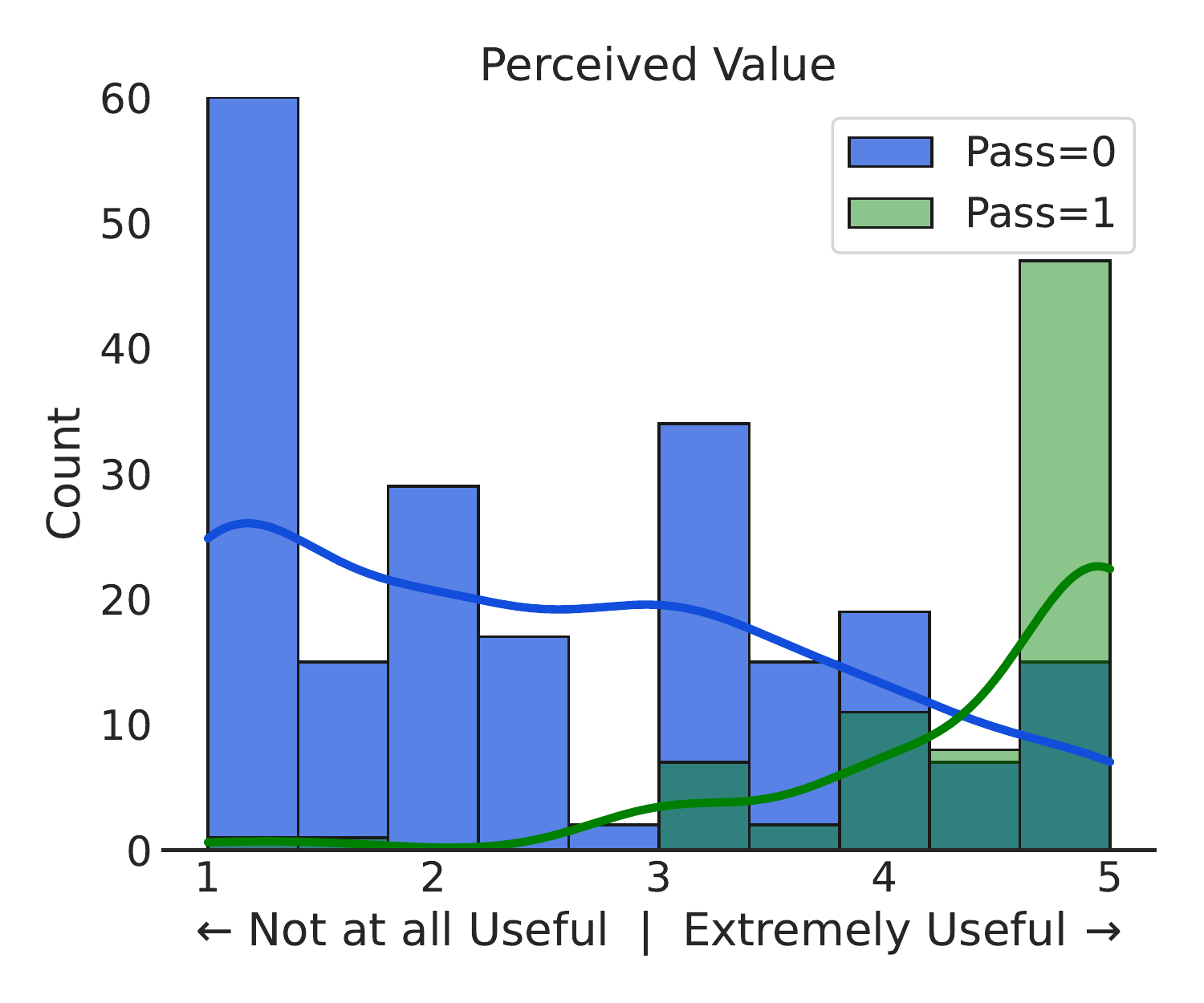}
  \caption{Distributions of \perValue judgments, in cases where generations pass unit tests ($\passOne=1$, green), and fail unit tests (\passOne=0, blue). When generations pass unit tests, they are likely to be judged as valuable ($\perValue \geq 3$). In fact, only 3\% of generations fall below this score. When functions fail unit tests, they are somewhat more likely to have lower \perValue scores, but 90 generations (42\%) are still considered ``somewhat useful'' by participants. The metric misses many high-value generations.
  \label{fig:valueacceffort}} 
\end{figure}

 \begin{figure}[t]
  \centering
    \includegraphics[width=\columnwidth]{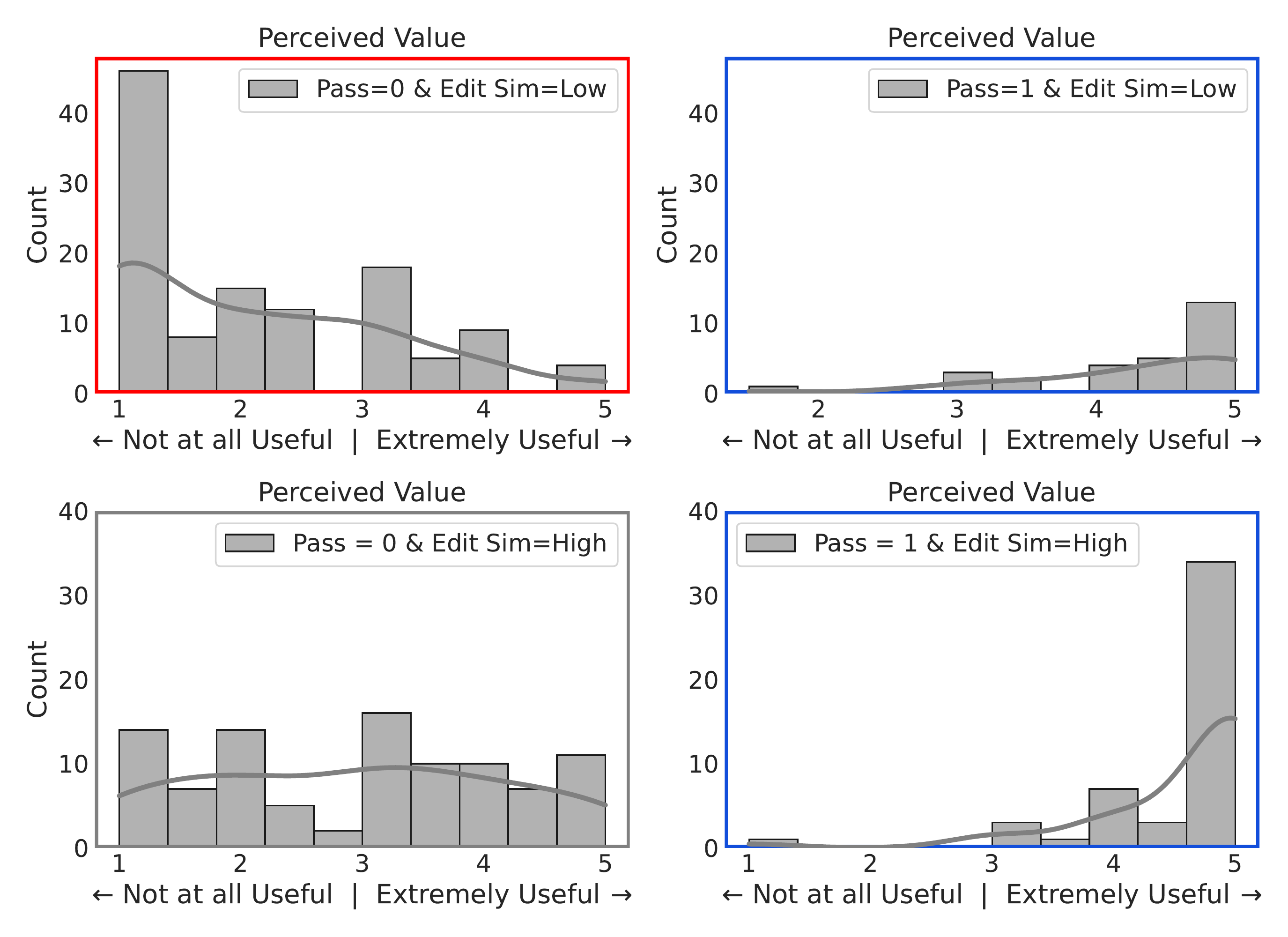}
 \caption{Distributions of \perValue judgments, in the four possible combinations of \passOne outcomes and low/high \editSim scores. The top and bottom rows indicate cases where \editSim falls below and above the 50\textsuperscript{th}-percentile, respectively. The left and right columns indicate cases where $\passOne=0$ and $\passOne=1$ respectively. When $\passOne=1$, generations are likely be high value (blue regions). When both $\passOne=0$ and \editSim=low, generations are likely to be low value (red region). The \perValue is more uniformly distributed in the remaining region where both $\passOne=0$ and \editSim=high. \label{fig:valuewrtpassedit} } 
\end{figure}

Our second finding confirms \textbf{that the metrics used in practice (\passOne, \editSim, and \bleu) are indeed positively correlated with \perValue}, but there are important differences (Fig. \ref{fig:studyheader}, bottom-left quadrant). As an example, \passOne shows the strongest association with \perAccuracy of the three metrics ($r = 0.66; p < 0.001$). This is unsurprising, given that \passOne is a direct measure of functional correctness. More surprising to us, however, is that \passOne is also the most strongly correlated metric to both \perEffort and \perValue ($r = 0.62; p < 0.001$, and $r = 0.62; p < 0.001$ respectively). This was unexpected since \editSim is a direct measure of the number of changes needed to correct a suggestion, and yet shows weaker correlation to \perEffort ($r = 0.48; p < 0.001$). With a correlation of $r = 0.36; p < 0.001$, \textbf{\bleu under-performs all other metrics}. Finally, given that none of the metrics correlate better than $r = 0.66$, there is significant opportunity to develop improved metrics.

\subsection{Code That Passes Unit Tests ($\passOne = 1$) is Extremely Likely to be High-Value}
Our third finding is that \textbf{when $\passOne = 1$ (i.e., when generations pass unit tests), we can be reasonably certain that they will be of high \perValue} (Figure \ref{fig:valueacceffort}). In fact, only 2 of 77 (3\%) generations that passed unit tests were found to have a \perValue scores less than 3. Recall, a \perValue score of 3 indicates that the participant found a suggestion to be at least ``somewhat useful.'' 

However, $\passOne = 0$ is less good at filtering low-value suggestions; Only 123 of the 213 (58\%) generations that failed unit tests scored lower than 3 on value. Stated differently, 90 generations (42\%) were found to be at least somewhat valuable.  This finding confirms existing qualitative work that while programmers value functionally correct code, they may still find considerable value in code that is not functionally correct \cite{weisz2021perfection}.

\subsection{Improved Metrics Through Combination} 
\label{sec:combinedmetric}
Upon further inspection, we realized that \editSim was itself a useful lens through which to understand how \perValue is distributed when unit tests are failed. Figure \ref{fig:valuewrtpassedit} shows a partitioning of results such that left and right columns correspond to ($\passOne = 0$) and ($\passOne = 1$) respectively. Top and bottom rows correspond to cases where the \editSim is below and above the 50\textsuperscript{th}-percentile respectively (referred to as \editSim=Low and \editSim=High). As before, we find that when ($\passOne = 1$), the \perValue tends to be high (blue outlined regions). However, we also find that \textbf{when a generation both fails the unit test and has low \editSim (i.e., \passOne $=0$; \editSim $=low$), it tends to be judged to have low \perValue} (red outlined region). Conversely, in the final region (\passOne $=0$; \editSim $=high$), \perValue is distributed more uniformly, and the signal is less clear. This strongly suggests that if human labeling is limited by budget, it may be worthwhile oversampling this region to maximally recover some of the missing \perValue signal.

This also suggests that there is an opportunity to combine metrics because $\passOne = 1$ is good at spotting high-value generations, while \passOne $=0$; \editSim $=high$ is good at spotting low-value generations. To investigate this further, we formally define a simple combined metric as follows:
\begin{equation*}
    \combined = \textit{min}(1.0, \passOne + \editSim )
\label{eq:combinedMetric}
\end{equation*}
Figure \ref{fig:studyheader}, row 7, shows some merit to this approach: The combined metric correlates better with human judgments of value ($r = 0.70; p < 0.001$) than \passOne ($r = 0.61; p < 0.001$) and \editSim ($r = 0.48; p < 0.001$) for \editSim. This is an extremely promising result, but was also only our first attempt at combining metrics. Future work is needed to explore other potential combinations.

\section{Discussion \& Future work}
\subsection{What Do Programmers Value?}
Much of the current research evaluating code generation models aims to approximate overall value via some notion of correctness \cite{chen2021evaluatingcodex,fried2022incoder,austin2021programgoogle,chowdhery2022palm,nijkamp2022conversationalcodegen,hendrycks2021measuring, kulal2019spoc}. Even research exploring similarity-based metrics have tended to validate these against some general notion of code quality (e.g., \citet{mathur-etal-2020-tangled} consider ``adequacy" while \citet{ren2020codebleu} consider ``good vs bad"). In this work, we aim to tease out distinct aspects of value to better understand how they contribute what programmers want from their AI-pair programmers. In this study, we examine the impact of correctness and effort. Our findings show that effort indeed matters to programmers. Accuracy also matters but, interestingly, our findings suggest that effort savings may be even more valuable to programmers than accuracy. 

In general, we take the position that value is a multidimensional theoretical construct \cite{THOMAS2022100476}. As such, while our findings showed effort as more valuable to programmers than accuracy, because both are still highly correlated with value, we recommend considering both when assessing the impact of human-AI pair programming. Moreover, there are likely many other properties of AI-pair programmers that developers find valuable \cite{forsgren2021the} and future work warrants investigating how these may also be captured in offline evaluations.

\subsection{How can Developers Approximate Value?}
Our results show that when developers have access to 
evaluation data containing high quality unit tests (as in HumanEval), generations that pass unit tests are highly likely to be valuable to programmers. This suggests that \passOne could be used as a reasonable filter in high precision scenarios (e.g., if an AI-pair programmer was tuned to limit distractions by only showing generations when they most likely to be valuable). 

That said, however, \passOne alone may miss a significant fraction of generations 
that programmers might 
find valuable. Our findings show that another offline metric -- \editSim can help overcome this issue when we combine it with \passOne according to Equation~\ref{eq:combinedMetric}.
This new metric is similar in spirit to {\em hinge-loss} in support vector machines.\footnote{\url{https://en.wikipdia.org/wiki/Hinge_loss}} In that setting, misclassifications are penalized based on their distance to the hyperplane decision boundary. Conversely, correct classifications all receive a fixed loss of 0, following the intuition that they don't become \emph{more correct} the further they move from the hyperplane. In our setting, we expect \perValue to increase as generations become more similar to a reference solution, but once it reaches functional correctness it doesn't become \emph{more correct} the closer it gets (syntactically) to the reference solution.

We emphasize, however, that metrics can have varying implications on model development decisions and therefore the choice of when or if to combine them is important. For example, when developers are seeking to make deployment decisions between models, selecting models that rank highest in terms of the overall value they may provide to programmers seems reasonable. In this case, the theoretical construct being approximated is perceived \perValue and our \combined metric is better at estimating this than \passOne or \editSim alone. However, when developers are diagnosing issues during model development (e.g., via error analyses)\textit{ we recommend that \passOne and \editSim be applied independently to get a clearer picture of model behavior} \cite{THOMAS2022100476} and to ensure appropriate mitigation strategies are used for different issues.
For example, \passOne failing on certain types of problems (e.g., recursive problems) or code blocks (e.g., conditional statements, error handling) may suggest additional data is needed in fine tuning. Whereas, \editSim failures may warrant new user interface techniques to help programmers focus attention to parts of the code most likely needing edits.

\subsection{Approximating Accuracy and Effort}
Our results show that programmers value both accuracy and effort savings when it comes to their AI pair programmers. We demonstrate that \passOne is a reasonable proxy for accuracy. Surprisingly, however, we found that \editSim is only moderately correlated with effort and in fact is less correlated with effort than \passOne. This is somewhat counter-intuitive since \editSim directly measures the number of characters that need to be changed to convert a generation to the reference solution \cite{svyatkovskiy2020intellicode,lu2021codexglue}.

This, along with our finding that programmers value effort reduction from their AI pair-programmers, suggests that an important area for future work is to experiment with alternative ways to operationalize effort for offline evaluation. This also, emphasizes the importance of validating that metrics faithfully capture the theoretical constructs they are trying to measure \cite{Jacobs_2021}.

\subsection{When Should Developers Use \editSim?}
Our findings show that \editSim is moderately correlated with \passOne. This is important because there are many situations where computing \passOne may be undesirable. For example, \passOne requires executing arbitrary generated code which can be resource intensive and may pose security risks \cite{chen2021evaluatingcodex}. \passOne and other functional evaluation metrics also require the availability of comprehensive, high-quality unit tests as well as language-specific test infrastructure, assumptions which may not hold in some evaluation scenarios (e.g., testing functions in the wild). Therefore, \textit{while we recommend \passOne when it is appropriate because it is more strongly correlated with value than \editSim, our findings suggest that \editSim may be a reasonable alternative when it is desirable to avoid limitations of functional evaluation}. 

Of course, limitations of similarity metrics should also be weighed against their benefits. For example, similarity metrics can fail when tasks have multiple syntactically divergent solutions - e.g. an algorithm may have an iterative vs recursive implementation with low token overlap, leading to noisy similarity metric. However, we  intuit that this scenario is relatively infrequent given the structured nature of programming languages and existing research on developer behaviour e.g., \citet{allamanis2018surveybigcode}  who mention that  \textit{developers prefer to write \cite{allamanis2014learning} and read code \cite{hellendoorn2015will} that is  conventional, idiomatic, and familiar, because it helps in understanding and maintaining software systems}. A convergence towards idiomatic solutions make it more likely the solutions and patterns learned by large language models of code coincide with ground truth solutions, limiting the scenario where generated code is syntactically different from but functionally equivalent to ground truth.

\section{Conclusion}

We studied how well two types of offline metrics for evaluating code generation models (i.e., functional correctness such as $pass@k$ based on unit tests and similarity-based metrics such as edit similarity) align with human judgements of value when used for human-AI pair programming. 
Our user study with 49 experienced programmers suggests that 
while programmers find functionally correct code generations valuable, the effort to edit and adapt generations also matters. Existing offline metrics show high correlation with human judgements of value, but there is room for improvement.
One reason is that while code that passes unit tests is very likely to be rated high-value,
code that fails unit tests is often still considered valuable by programmers. Based on this observation, we propose a combined offline metric inspired by hinge-loss in support vector machines that allows for partial credit by combining strengths of functional correctness and similarity-based metrics. Our analysis shows that this combined metric aligns better with human judgements of value in code generations than functional correctness or similarity alone. 
Overall our work highlights the importance of validating that offline metrics in AI capture what people value and that human-centered metrics, inspired by what people value, can provide better estimates of what people want from their AI-pair programmers.

 \newpage 



\newpage

\section*{Limitations}  
In this work, we focused on problems posed in the hand-crafted  HumanEval dataset \cite{chen2021evaluatingcodex}.  A potential pitfall of a curated dataset such as HumanEval is that the results \textit{may} not generalize to real-world scenarios where developers often deal with more complex problems and code bases (e.g, code with multiple dependencies across multiple files). To address this limitation, we originally explored the use of datasets mined from GitHub. However, our experiments indicated memorization issues (e.g., verbatim generation of solutions to niche problem), potentially due to the sample code already being included in the model training set\cite{lee2021deduplicating}. In practice, high quality code deduplication required to avoid this specific limitation is challenging. Work by \citet{allamanis2019adverse} find that the impact of duplicate code can be severe, sometimes inflating model performance scores by up to 100\%. Furthermore, in our early pilot tests, functions extracted in the wild were found to contain insufficient context (e.g. absence of docstring) for even expert human annotators and isolating functional tests is challenging without heavy curation. Further research is therefore needed to understand how our findings might generalize to a wider variety of deployment settings as well as research on designing diverse evaluation datasets.
In addition, future work may also explore the impact of problem difficulty on the observed results in our study.   

\section*{Ethics Statement}
While our study informs current practices in evaluating code generation models, we acknowledge that measures of value can differ across multiple demographics with impact on productivity. For our experiments (approved by an internal IRB board), we generate code snippets based on a publicly available dataset, using publicly available models that are annotated by experienced developers. These choices make our work readily reproducible. We also developed a library that implements multiple metrics for bench marking code generation models which we will make available as an open source library (MIT license) at the time of publication.



\bibliography{anthology,paper}
\bibliographystyle{acl_natbib}

%
%

\end{document}